\documentclass[multphys,vecphys]{svmult}

\usepackage{makeidx}			
\usepackage{graphicx}			
\usepackage{multicol}			
\usepackage[bottom]{footmisc}		

\makeindex				

\newcommand{\lapprox}{\,\rlap{\lower 2.5pt 
				\hbox{$\sim$}}\raise 1.5pt\hbox{$<$}\,}
\newcommand{\kms}{km~s$^{-1}$}                  
\newcommand{\msun}{\mbox{$M_{\odot}$}}          

\begin{document}

\title*{Globular Cluster Formation in Mergers}
\author{Fran\c cois Schweizer}
\institute{Carnegie Observatories, 813 Santa Barbara Street, Pasadena,
CA 91101, USA
\texttt{schweizer@ociw.edu}}

\maketitle

\begin{abstract}
Mergers of gas-rich galaxies lead to gravitationally driven increases in
gas pressure that can trigger intense bursts of star and cluster formation.
Although star formation itself is clustered, most newborn stellar
aggregates are unbound associations and disperse.
Gravitationally bound star clusters that survive for at least 10--20 internal
crossing times ($\sim$20--40 Myr) are relatively rare and seem to contain
$<$10\% of all stars formed in the starbursts.
The most massive young globular clusters formed in present-day mergers
exceed $\omega$ Cen by an order of magnitude in mass, yet appear to have
normal stellar initial mass functions.

In the local universe, recent remnants of major gas-rich disk mergers
appear as protoelliptical galaxies with subpopulations of typically
10$^2$--10$^3$ young metal-rich globular clusters in their halos.
The evidence is now strong that these ``second-generation'' globular
clusters formed from giant molecular clouds (GMC) in the merging disks,
squeezed into collapse by large-scale shocks and high gas pressure rather
than by high-velocity cloud--cloud collisions.
Similarly, first-generation metal-poor globular clusters may have formed
during cosmological reionization from low-metallicity GMCs squeezed by
the universal reionization pressure.
\end{abstract}

\section{On the Nature of Young Globular Clusters}
\label{sec1}

When studying the myriads of point-like luminous sources brighter than
any individual star on \emph{HST} images of \emph{ongoing} mergers
(e.g., NGC 4038/39, NGC 3256), one would like to know which ones---or
at least what fraction---will survive as globular clusters (GC).
Yet, it is very difficult to distinguish gravitationally bound young
star clusters from unbound OB associations or even spurious asterisms.
As it turns out, the adopted operational definition for ``cluster''
may determine the answers to the scientific questions we ask
about these objects.

Modern astronomical dictionaries universally include in their definition
of ``star cluster'' (open or globular) the requirement that it be
\emph{gravitationally bound}, thus distinguishing it from any looser,
expanding ``stellar association'' (e.g., \cite{hopk76,ridp97}).
As I explain in Sect.~\ref{sec2} below, I believe that our present
inability to make this distinction for many stellar aggregates younger
than 10--20~$t_{\rm cr}$ (internal crossing times) in ongoing mergers leads
to a notion of ``infant mortality'' that is seriously exaggerated.

In recent merger \emph{remnants}, where the merger-induced starburst has
subsided (e.g., NGC 3921, NGC 7252), the definition of a
\emph{young globular cluster (YGC)}
is more easy and secure. 
Any young compact stellar aggregate older than 10--20 $t_{\rm cr}$
($\sim$20--40 Myr), more massive than a few $10^4 M_{\odot}$, and with
a half-light radius $R_{\rm eff}$ comparable to that
of a typical Milky-Way globular (say, $R_{\rm eff} \lapprox 10$~pc) is
most likely gravitationally bound and, hence, a YGC. 
It is the size requirement that places stringent upper limits on
any possible expansion velocity ($\lapprox$0.2--0.5 \kms) and thus
guarantees that the cluster is gravitationally bound.

An important result to emerge from recent \emph{HST} and follow-up studies
of YGCs concerns their masses.
These masses do not only cover the full range observed in old Milky-Way
GCs ($\sim$10$^4$ -- $5\times 10^6 \msun$), but also extend to nearly
$10^8 \msun$ or $\sim$20$\times$ the mass of $\omega$~Cen at the high-mass
end.
The most massive YGCs are invariably found in remnants of gas-rich major
mergers such as NGC 7252 \cite{ss98,mara04}, NGC 1316 \cite{bast06}, and
NGC 5128 \cite{mart04}.
Interestingly, dynamical masses determined from velocity dispersions
agree well with photometric masses based on cluster-evolution models
with normal (e.g., Salpeter, Kroupa, or Chabrier) initial mass
functions (IMFs).
Therefore, some earlier worries that YGCs formed in mergers may have
highly unusual stellar IMFs (e.g., \cite{brod98}) seem now unfounded.

Relatively little work has been done so far on the brightness profiles
and detailed structural parameters (core and tidal radii) of YGCs in
mergers.
Yet, the subject looks promising.
Radial profiles of selected YGCs in NGC 4038 suggest that the initial
power-law envelopes of YGCs may be tidally stripped within the first
few 100 Myr, while the core radii may grow \cite{whit99}.
Correlations between core radius and cluster age are known to exist for
the young cluster populations of the Magellanic Clouds (e.g., \cite{mack03})
and deserve further study via the rich cluster populations of ongoing
mergers and merger remnants.

\section{Formation and Early Evolution}
\label{sec2}

Star clusters form in giant molecular clouds (GMC), where optical
extinction can be very significant.
Hence the question arises what fraction of all young clusters ``optical''
surveys made with \emph{HST} ($0.3\lapprox\lambda\lapprox1.0\,\mu$)
may miss.

This question has been addressed by Whitmore \& Zhang \cite{wz02} for the 
``Overlap Region'' of NGC 4038/39, which is known to harbor some of the
most IR-luminous young clusters, yet appears heavily extincted at optical
wavelengths and brightly emitting at 8$\mu$ \cite{wang04}.
A comparison between optical clusters and strong thermal radio sources
shows that 85\% of the latter have optical counterparts, whence even in
this extreme region only $\sim$15\% of all clusters have been missed by
\emph{HST} surveys \cite{wz02}. 
Measured cluster extinctions lie in the range
$0.5\lapprox A_V\lapprox 7.6$ mag and diminish to $A_V\lapprox 1.0$ mag
for clusters 6 Myr and older.
This suggests that cluster winds disperse most of the natal gas rapidly,
and that optically-derived luminosity functions for clusters older
than $\sim$6 Myr should not be too incomplete.

\subsection{Cluster Luminosity Functions}
\label{sec21}

To first order, the luminosity functions (LF) of young-cluster systems
in merger galaxies are well approximated by a power law of the form\ \ 
$\Phi(L) dL \propto L^{-\alpha} dL$\ \ with
$1.7\lapprox \alpha\lapprox 2.1$\ \ \cite{ws95,meur95,whit03}.
The similarities between this power law and the power-law mass function
of GMCs, including the similar observed mass ranges, strongly suggest
that young clusters form from GMCs suddenly squeezed by a rapid increase
in the pressure of the surrounding gas \cite{jog92,hapu94,elme97} (see
also Sect.~\ref{sec23}).

\begin{figure}
  \centering
  \includegraphics[width=11.6truecm]{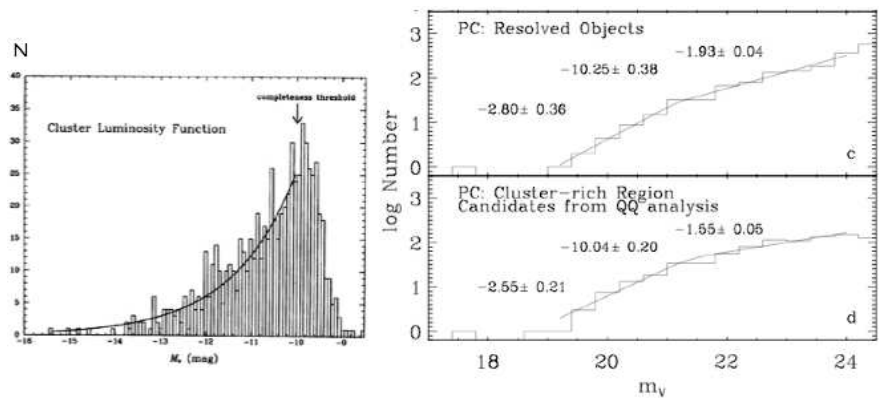}
  \caption{Luminosity functions for candidate young star clusters in
	NGC 4038/39 from \emph{HST} observations with
	\textit{(left)} WFC1 \cite{ws95} and
	\textit{(right)} WFPC2 \cite{whit99}.}
  \label{fig1}       				
\end{figure}

Deep \emph{HST} observations of mergers with rich cluster systems suggest
that the cluster LFs may have a break (``knee'') whose position varies
from merger to merger (NGC 4038/39 \cite{whit99}; NGC 3256 \cite{zepf99};
M51 \cite{giel06}).
Figure~\ref{fig1} displays for NGC 4038/39 both the original cluster LF
\cite{ws95} and two versions of the deeper LF \cite{whit99} showing a
break around $M_V = -10.0$ to $-$10.3.
The interpretation of these breaks is presently controversial.
Either the breaks reflect brightness-limited-selection effects (Whitmore
et al., in prep.) or they may indicate a maximum cluster mass
\cite{giel06}.
In the latter case, the measured LF breaks in the above three mergers
would seem to suggest that the maximum mass increases with the vehemence
of the merger, presumably indicating that under increased gas pressure
GMCs coagulate into more massive aggregates.

\subsection{Star-Cluster Formation vs Clustered Star Formation}
\label{sec22}

The \emph{age distribution} of young clusters in NGC 4038/39 has recently
been derived for two mass-limited subsamples defined by
$M > 3\times 10^4 \msun$ and $M > 2\times 10^5 \msun$ \cite{fall05}.
The masses themselves are estimates based on \emph{HST} photometry in
$U\!BV\!I$ and H$\alpha$ plus Bruzual-Charlot \cite{bc03} cluster evolution
models.
The number distributions for both subsamples decline steeply with age
$\tau$, approximately as $dN/d\tau \propto \tau^{-1}$.
Thus, it would seem that $\sim$90\% of all clusters disrupt during each
age decade.
The median age of the clusters is a mere $\sim$10$^7$ yr, which Fall et
al.\ interpret as evidence for rapid disruption, dubbed ``infant mortality.''
These authors guess that ``very likely ... most of the young clusters
are not gravitationally bound and were disrupted near the times they
formed by the energy and momentum input from young stars to the ISM of
the protoclusters.''

In my opinion, it is unfortunate that this loose, non-astronomical use
of the word ``cluster'' may reinforce an increasingly popular view that
most stars form in clusters.
By the traditional astronomical definition of star clusters as
gravitationally bound aggregates, most of the objects tallied by Fall
et al.\ in The Antennae are not clusters, but likely young stellar
associations.
It seems to me in much better accord with a rich body of astronomical
evidence gathered during the past 50 years to state that---although
\emph{star formation is clearly clustered}---even in mergers
gravitationally bound clusters (open and globular) form relatively rarely
and \emph{contain $<$10\% of all newly-formed stars}.

I believe that only with such careful distinction can we hope to study
the true disruptive effects that affect any gravitationally bound star
cluster over time, including mass loss due to stellar evolution and
evaporation by two-body relaxation and gravitational shocks.

Further reason for caution is provided by the recent discovery that even
in nearby M31, four of six claimed YGCs have turned out to be spurious
asterisms when studied with adaptive optics \cite{cohe05}.
Clearly, there is considerable danger in calling all luminous point-like
(at \emph{HST} resolution) sources in the distant NGC 4038/39 young
``clusters''!

\subsection{Shocks and High Pressure}
\label{sec23}

Shocks and high pressure have long been suggested to be the main drivers
of GC formation in gas-rich mergers and responsible for the increased
specific frequency $S_N$ of GCs observed in descendent elliptical galaxies
\cite{schw87,jog92,az92}.

Much new evidence supports this hypothesis.
\emph{Chandra} X-ray observations of the hot ISM in merger-induced
starbursts, and especially in NGC 4038/39 \cite{fabb04}, show
that the pressure in the hot, 10$^6$--10$^7$K ISM of a merger can
exceed 10$^{-10}$ dyn cm$^{-2}$ and is typically 10--100 times higher
than it is in the hot ISM of our local Galactic neighborhood (e.g.,
\cite{bald06,veil05}).
Thus GMCs in mergers do indeed experience strongly increased pressure
from the surrounding gas.

The principal source of general pressure increase are gravitational
torques between the gas and stellar bars, which tend to brake the gas
and lead to rapid inflows and density increases (e.g., \cite{bh96,mh96}).

What has become clearer only recently is how much accompanying
\emph{shocks} may affect the spatial distribution of star and cluster
formation.
As Barnes \cite{barn04} shows via numerical simulations,
star-formation recipes that include not only the gas density (i.e.,
Schmidt--Kennicut laws), but also the local rate of energy dissipation
in shocks, lead to spatially more extended star and cluster formation that
tends to occur earlier during the merger.
A model with mainly shock-induced star formation for The Mice (NGC 4676)
leads to significantly better agreement with the observations of H$\;$II
regions and young clusters than one with only density-dependent star
formation.
Shock-induced star formation may also explain why cluster formation is
already so vehement and wide-spread in The Antennae, where the two
disks---currently on their second approach---are still relatively intact.

\begin{figure}
  \centering
  \includegraphics[width=11.6truecm]{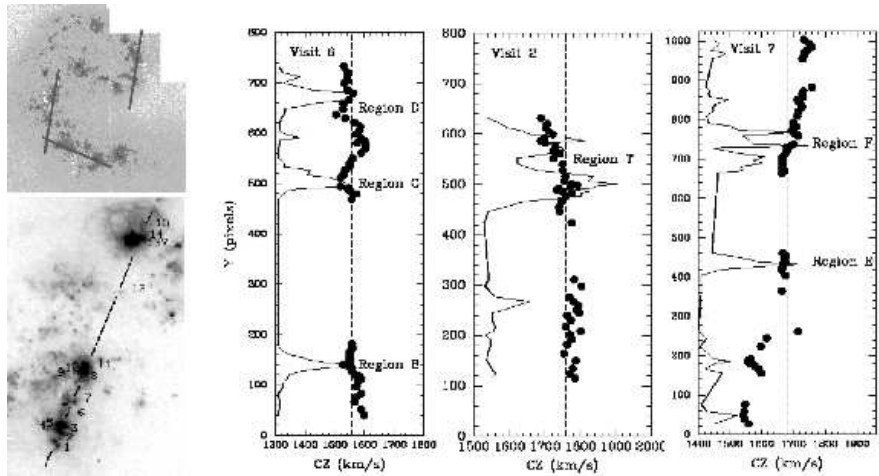}
  \caption{Radial velocities of young clusters in NGC 4038/39, measured
	with \emph{HST}/STIS (at H$\alpha$) along three lines crossing
	7 major regions, each with many clusters.  The three slit positions
	are shown at \textit{upper left}, while \textit{lower left panel}
	shows slit position across regions D, C, and B in more detail.
	After gradient subtraction, the cluster-to-cluster velocity
	dispersion is \,$<$10--12 \kms\ \cite{whit05}.}
  \label{fig2}       				
\end{figure}

Are the shocks in mergers generated by high-velocity, 50--100 \kms\
cloud--cloud collisions \cite{kuma93} or more by large-scale gas
motions?
A high-resolution study with \emph{HST}/STIS of the radial velocities of
many dozens of young clusters in 7 regions of The Antennae shows that the
average cluster-to-cluster radial-velocity dispersion is\ \ 
$\sigma_{\rm v,cl} < 10$--12 \kms\ \cite{whit05}, as
illustrated in Fig.~\ref{fig2}.
This relatively low velocity dispersion argues strongly against
high-velocity cloud--cloud collisions and in favor of the general pressure
increase being what triggers GMCs into forming clusters \cite{jog92,elme97}.

\section{Young Metal-Rich Halo Globulars}
\label{sec3}

There are several advantages to studying YGCs in relatively recent,
about 0.3--3 Gyr old  merger remnants:
(1) Dust obscuration is much less of a problem than in ongoing mergers.
(2) Most point-like luminous sources in such remnants are true GCs,
since time has acted to separate the wheat from the chaff (= expanding
associations), and clusters are now typically $>$100 Myr or
$>$25--50 $t_{\rm cr}$ old.
And (3), the remnants themselves appear to be evolving into bona fide
early-type galaxies.
Therefore, YGCs formed during the mergers can provide key evidence on
processes that must have shaped GC populations in older E and S0 galaxies
as well.

\emph{HST} studies of recent merger remnants such as NGC 3921 \cite{schw96},
NGC 7252 \cite{mill97}, and NGC 3597 \cite{carl99} show that these galaxies
typically host about 10$^2$--10$^3$ point-like sources that appear to be
mostly \emph{young} GCs ($\lapprox$1 Gyr old).
(It is not that there are no old GCs in these relatively distant remnants,
only that the YGCs are much brighter and more easily studied.)
Age-dating based both on broad-band photometry and spectroscopy shows that
the majority of these YGCs formed in relatively short, 100--200 Myr time
spans during the mergers.
The YGCs appear strongly concentrated toward their hosts' centers, half
of them lying typically within $\lapprox$5 kpc from the nucleus.

The few spectroscopic studies that have so far been made of such YGCs
invariably show them to be of approximately solar metallicity:
$[Z] = 0.0\pm 0.1$ in NGC 7252 \cite{ss98}, $0.0\pm 0.5$ in NGC 3921
\cite{schw04}, and---for the intermediate-age,
$\sim$3--5 Gyr old GCs in more advanced remnants---$[Z] = 0.0\pm 0.15$
in NGC 1316 \cite{goud01} and $-0.1\pm 0.2$ in NGC 5128 \cite{peng04}.

Such near-solar metallicities in recently formed GCs are, of course, not
unexpected and might not seem worth emphasizing, were it not for the fact
that the YGCs with these metallicities all show \emph{halo} kinematics
(see refs.\ above).
Therefore, the inevitable conclusion is that major mergers of gas-rich
disk galaxies produce \emph{young metal-rich halo GCs}.
The existence of significant populations of such clusters in merger remnants
ranging from $\sim$0.5 Gyr to 4--5 Gyr in age, together with observational
and theoretical evidence that the remnants themselves are young to
intermediate-age ellipticals, provides a strong link to the old metal-rich
GC populations observed in virtually all E and many S0 galaxies (see
\cite{schw03} and Goudfrooij's contribution in this volume for further
details).

\section{Implications for Old Metal-Poor Globular Clusters}
\label{sec4}

Perhaps the main result from studies of GC formation in mergers is that
the process is driven by strong pressure increases that squeeze GMCs
into rapid cluster formation.
Observations show that the pressures in the ISM can exceed
10$^{-10}$ dyn cm$^{-2}$ already early on in a merger (Sect.~\ref{sec23}),
while simulations of gas-rich mergers demonstrate that most of the pressure
increase is driven gravitationally \cite{bh96,mh96,barn04}.

These facts beg the question whether some nearly universal pressure
increase may have caused the formation of the old metal-poor GCs that
are so omnipresent in all types of galaxies and environments.

Cen \cite{cen01} points out that the cosmological reionization at
$z \approx 15$--7 may have provided just such a universal pressure increase.
Ionization fronts driven by the external radiation field may have
generated inward convergent shocks in gas-rich sub-galactic halos,
which in turn triggered GMCs into forming clusters.
If so, the formation of metal-poor GCs from early GMCs in many
of these halos may have been nearly synchronous.

If Cen's hypothesis is correct, most GCs in the universe may have
formed from shocked GMCs.
The first-generation GCs formed near-simultaneously from low-metallicity
GMCs shocked by the pressure increase accompanying cosmological
reionization. 
Later-generation (``second-generation'') GCs formed during subsequent
galaxy mergers from metal-enriched GMCs present in the merging components
and shocked by the rapid, gravitationally-driven pressure increases of
the mergers.
Major disk mergers, some of which occur to the present time, led to
elliptical remnants with a mixture of first- and second-generation
GCs that can still be traced  by their bimodal color distributions.
Finally, a minority of second-generation GCs seem to form sporadically
from occasional pressure increases in calmer environments, such as in
interacting irregulars and barred spirals.

\section{Conclusions}
\label{sec5}

During mergers, increased gas pressure leads to much \emph{apparent}
cluster formation, but most of the stellar aggregates are unbound and
disperse.
Gravitationally bound globular and open clusters are relatively \emph{rare}
and seem to contain $<$10\% of all stars formed in the starbursts.

Major gas-rich mergers form not only E and S0 galaxies, but also their
metal-rich ``second-generation'' GCs.
Specifically, in the local universe young remnants of major such mergers
appear as protoelliptical galaxies with subpopulations of young
metal-rich halo GCs (NGC 3921, NGC 7252; later NGC 1316, NGC 5128).
The evidence is now strong that these second-generation GCs form from
giant molecular clouds in the merging disks, squeezed into collapse
by large-scale shocks and high gas pressure rather than by high-velocity
cloud--cloud collisions.

Similarly, first-generation metal-poor GCs may have formed during
cosmological reionization from low-metallicity giant molecular clouds
squeezed by the reionization pressure.
\\ \\
\textbf{Acknowledgement.} I thank Brad Whitmore for his permission to
reproduce some figures.


\printindex
\end{document}